\documentclass[pdftex,twocolumn,epjc3]{svjour3}          
\RequirePackage[T1]{fontenc}
\smartqed  
\RequirePackage{graphicx}
\RequirePackage{mathptmx}      
\RequirePackage{flushend}
\RequirePackage[numbers,sort&compress]{natbib}
\RequirePackage[colorlinks,citecolor=blue,urlcolor=blue,linkcolor=blue]{hyperref}

\journalname{Eur. Phys. J. C}
\usepackage{graphicx}
\usepackage{amssymb}
\usepackage{amsmath}
\usepackage{color}
\usepackage{hyperref}
 \usepackage{tabu}

\def\barray{\begin{array}}
\def\earray{\end{array}}
\def\be{\begin{equation}}
\def\ee{\end{equation}}
\def\ben{\begin{equation} \nonumber}
\def\een{\end{equation}}
\def\ban{\begin{eqnarray*}}
\def\ean{\end{eqnarray*}}
\def\ba{\begin{eqnarray}}
\def\ea{\end{eqnarray}}

\def\({\left(}
\def\){\right)}

\graphicspath{{./fig/}}

\begin{document}

\title{Intermediate Class of Warm Pseudoscalar Inflation}
\author{Saeid Ebrahimi \thanksref{e,addr},
Vahid  Kamali\thanksref{e1,addr},
Asma Alaei\thanksref{e2,addr}}

\thankstext{e}{E-mail: saeid.ebrahimi1365@gmail.com}
\thankstext{e1}{E-mail: vkamali@basu.ac.ir}
\thankstext{e2}{E-mail: asmaalaii@yahoo.com}

\institute{Department of Physics, Bu-Ali Sina (Avicenna) University, Hamedan 65178, 016016, Iran\label{addr}
}
\date{Received: date / Accepted: date}
\maketitle
\begin{abstract}
High dissipative regime of warm pseudoscalar inflation model \cite{Kamali:2019ppi} with an approximately constant value of dissipation parameter $Q$ is studied. { Intermediate solution of the scale-factor related to the accelerated expansion of the  Universe which is rolled out by observational data in the context of standard (cold) model of inflation is used.} There is a region of free parameters phase-space of the model which is interestingly compatible with recent observational data. It is discussed that the model is also compatible with the swampland criteria in a broad range of parameters phase-space and TCC in a limited area of parameters. 
   
\end{abstract}
\maketitle
\section{Introduction:}
{ The inflation model is extensively studied as a standard paradigm of early time cosmology \cite{Starobinsky:1979ty,Starobinsky:1980tf,Starobinsky:1980te, Sato:1980yn, Guth:1980zm, Albrecht:1982wi}. Outstanding concerns of the Hot Big Bang model, e.g. horizon, flatness, and monopole, are resolved in this context. A brilliant achievement of this model is presented in the context of perturbation theory. The (scalar) field theory responsible for background accelerated expansion provides a mechanism to explain the seeds of large-scale structure (LSS) formation in terms of quantum fluctuations of the field around the homogeneous and isotropic background  \cite{Mukhanov:1981xt,Hawking:1982cz,Guth:1982ec,Starobinsky:1982ee,Bardeen:1983qw}. The standard model of inflation is proposed by a single scalar field theory with a nearly flat potential which is followed by a reheating epoch \cite{Kofman:1997yn,Allahverdi:2010xz,Shtanov:1994ce,Kofman:1994rk,Traschen:1990sw}. The idea of early time accelerated expansion, in term of the effective field theories that seem to be in the string-landscape, is doubted in the literature \cite{Ooguri:2006in,Obied:2018sgi,Brennan:2017rbf, Palti:2019pca}.}
 It was speculated that the effective field theory related to standard single field inflation is not in the landscape of string theory as a preferred scenario for quantum gravity \cite{Obied:2018sgi, Agrawal:2018own}. On the other hand, the energy scale and duration of the standard inflation are constrained by Trans-Planckian censorship conjecture TCC \cite{Bedroya:2019snp} which made the highest limitation on the tensor-to-scalar ratio far from the accuracy of cosmological experiments. This conjecture is a cosmological extension of cosmic censorship conjecture by Penrose \cite{Penrose:1969pc,Brandenberger:2019jbs}. 
  There are attempts in the literature  to find the ways that probably resolve these theoretical constraints \cite{ Heisenberg:2018yae, Kamali:2019gzr, Das:2018rpg, Motaharfar:2018zyb, Berera:2019zdd, Kamali:2019hgv,Laliberte:2019sqc,Dhuria:2019oyf,Torabian:2019zms,Li:2019ipk,Cai:2019hge,Brahma:2019unn, Das:2019hto, Kamali:2019xnt, Lin:2019pmj,Kadota:2019dol, Mizuno:2019bxy, Kamali:2019wdh,  Berera:2020iyn, Berera:2020dvn}. 
Warm inflation as an alternative for the standard (cold) inflation is introduced as a more than one field model of inflation \cite{Berera:1995ie}.
There is an interaction between the inflaton field as quanta of the scalar field and light (radiation) fields during the slow-roll epoch which smoothly connects the inflation era to radiation dominated era. It has been discussed that some scenarios of high dissipative warm inflation are in the landscape of string theory and compatible with observational data \cite{Motaharfar:2018zyb}. Intermediate model of inflation \cite{Barrow:1990td,Barrow:1993zq,Barrow:2006dh} is rolled out by observational data in the context of cold inflation but agrees with latest data in the context of warm inflation in some cases \cite{delCampo:2009xi, Setare:2012fg, Setare:2013dd, Setare:2013kja, Setare:2014oka, Kamali:2015yya, Kamali:2016frd}. {In this work, we will study a model of warm inflation which introduces the standard Chern-Simons interaction between pseudoscalar field as inflaton and SU(2) gauge fields as radiation part \cite{Kamali:2019ppi}, by using an intermediate solution of Universe accelerated expansion \cite{Barrow:1990td,Barrow:1993zq,Barrow:2006dh}. It was discussed that the dissipation parameter of warm pseudoscalar  inflation (WPSI) is approximately constant \cite{Kamali:2019ppi}.} We have found a region of free parameters phase space of the intermediate WPSI compatible with observational data as well as quantum gravity (swampland and TCC). We can simply find a region of parameters for the model out of swampland. There is a limited region of the parameters compatible with TCC and observation.

\section{Intermediate warm inflation}
The intermediate form of the scale factor \cite{Barrow:1990td,Barrow:1993zq,Barrow:2006dh}:
\begin{eqnarray}\label{scale}
a(t)=a_0\exp(At^f)
\end{eqnarray}
was introduced as a cosmological solution of an expanding Universe where the expansion is faster than power-low inflation ($a(t)=a_0t^A$) and slower than standard de Sitter expansion with the scale factor $a(t)=a_0\exp(Ht)$ ($H$ is  Hubble constant). This form of the scale factor is useful in studying the models of inflation \cite{delCampo:2009xi, Setare:2012fg, Setare:2013dd, Setare:2013kja, Setare:2014oka, Kamali:2015yya, Kamali:2016frd}.
Generally, to study and parameterize a model of inflation, we have to use a special form of the potential with some constant parameters that are supposed to be constrained by observational data. On the other hand, there is another nearly equivalent way where the form of scale factor, with some constant parameters, is used, and its  phase space of free parameters is constrained by observational data \cite{Barrow:1990td,Barrow:1993zq,Barrow:2006dh}.
In this note, we study our model by using the scale factor (\ref{scale}) which has two free parameters $(f, A)$. 
In the warm scenario of inflation where the particle production is important during the slow-roll epoch, there are some modifications in the evolution of the inflaton, background, and perturbation parts, in a thermal bath with temperature $T$. At the background level we can simply split the continuity equation into two parts with a phenomenological form of interaction:

\begin{eqnarray}\label{contin}
\dot{\rho}_{\phi}+3H(\rho_{\phi}+P_{\phi})=-\Upsilon\dot{\phi}^2
\\
\nonumber
\dot{\rho}_{\gamma}+3H(\rho_{\gamma}+P_{\gamma})=\Upsilon\dot{\phi}^2
\end{eqnarray}
The first part provides a modified form of the inflaton evolution:
\begin{eqnarray}
\ddot{\phi}+3H(1+Q)\dot{\phi}+\frac{dV}{d\phi}=0
\end{eqnarray} 
where $Q=\frac{\Upsilon}{3H}$ is introduced as a new dissipation coefficient beyond the Universe expansion dissipation $H$. 
$Q$ is an extra parameter in the context of warm inflation that is supposed to be constrained by observation data as well as quantum gravity. Therefore our model has a three-dimensional phase space of free parameters  $(f, A, Q)$ which will be studied in the next section.  
 In a high dissipative regime, $Q\gg 1$, dissipation term has an important role in the evolution of the inflaton in the background and linear perturbation parts which can be compared with observational data. { Dissipation parameter $Q$ is approximately constant in the context of WPSI \cite{Kamali:2019ppi}, where the interaction term between axion inflaton field and SU(2) gauge-field (as a radiation part) is presented by standard Chern-Simons term}{\footnote{{ We note that there are developments in this context where the gauge fields are not just SU(2) and  dissipation parameter is proportional to $T^3$  \cite{Berghaus:2019whh,Goswami:2019ehb, Das:2020xmh}. We have also checked this form of dissipation using intermediate scale factor (\ref{scale}) but the spectral index is out of the preferred region of CMB observation  ($n_s-1=-\frac{1.5f+27}{4(f(N-1)+1})$) for all possible values of parameter $0<f<1$. }  } } :
\begin{eqnarray}\label{int}
\mathcal{L}_{int}=\frac{\phi}{8M}F^a_{\mu\nu}\tilde{F}_a^{\mu\nu}~~~~~~~~~~~~~~~~\\
\nonumber
F^a_{\mu\nu}=\frac{1}{ig}[D^a_{\mu}, D^a_{\nu}]~~~~~\tilde{F}_a^{\mu\nu}=\frac{1}{2}\epsilon^{\mu\nu\rho\sigma}F_{a\mu\nu}\\
\nonumber
D^a_{\mu}=\partial_{\mu}-igA_{\mu}J^a~~~~~~~~Tr[J_a,J_b]=\frac{1}{2}\delta_{ab}
\end{eqnarray}
{Using interaction term (\ref{int}), we can derive the dissipation parameter $Q=\frac{g\psi^2}{2M H}$ (where $\psi$ relates to the gauge fields) for warm inflation which is approximately constant \cite{Kamali:2019ppi}.
In the linear perturbation part the modification by interacting term (\ref{int}) introduces a new version of the scalar power-spectrum:}
\begin{eqnarray}\label{Scalar power-spectrum}
\Delta_R^2=\frac{V(1+Q)}{24\pi^2M_p^2\epsilon_{\phi}}(1+2n+\frac{2\sqrt{3}Q}{\sqrt{3+4\pi Q}}\frac{T}{H})
\end{eqnarray}
which will be used in our study. In warm inflation there is non-negligible radiation energy density which is subdominant during slow-roll inflation. The first Friedmann equation is modified as:
\begin{eqnarray}\label{Fried}
H^2=\frac{1}{3M_p^2}(\rho_{\phi}+\rho_{\gamma})
\end{eqnarray}
where the evolution of $\rho_{\gamma}=C_{\gamma}T^4$ is presented by the second equation of (\ref{contin}). Near the horizon crossing point, the evolution of the universe is inflaton dominated $\rho_{\gamma}<\rho_{\phi}$ which is the condition that we consider to obtain the background parameters of the model at the first step and the perturbation part in the second step. {Using Eqs.(\ref{scale}), (\ref{contin}) and Eq.(\ref{Fried}) we present the evolution of the scalar field during the cosmic expansion \cite{Rendall:2005if, delCampo:2009xi, delCampo:2009ma, Herrera:2010yg, Herrera:2011zz, Herrera:2013rra}:
\begin{eqnarray}\label{scalar}
\dot{\phi}^2=-\frac{2M_p^2\dot{H}}{1+Q}~~~~~~~~\phi=\phi_0+\sqrt{\frac{8 M_p^2A(1-f)}{fQ}}t^{\frac{f}{2}}
\end{eqnarray}
where $\phi_0$ is the constant of integration that can be neglected without loos the generality. 
Cosmic time of the model is simply presented as a function of the scalar field (\ref{scalar}) 
\begin{eqnarray}\label{Cosmic time}
t=(\frac{fQ}{8M_p^2 A(1-f)})^{\frac{1}{f}}\phi^{\frac{2}{f}}
\end{eqnarray}
which is an important equation in our discussion. }
Using Eqs.(\ref{Fried}) and (\ref{Cosmic time}), we can find the potential
in term of the scalar field:

\begin{eqnarray}\label{Potential}
V(\phi)=9M_p^2f^2A^2[\frac{fQ}{8M_p^2 A(1-f)}]^{\frac{2(f-1)}{f}}\phi^{\frac{-4(1-f)}{f}}
\end{eqnarray}
{where $H=fAt^{f-1}$ (\ref{scale})}. The slow-roll parameters are defined as {(Eqs.(\ref{scale}) and (\ref{Cosmic time}) have been used)}:
\begin{eqnarray}\label{slow-roll}
\epsilon=-\frac{\dot{H}}{H^2}=\frac{8 M_p^2(1-f)^2}{f^2 Q}\frac{1}{\phi^2}\\
\nonumber
\eta=-\frac{\ddot{H}}{H\dot{H}}=\frac{8 M_p^2(1-f)(2-f)}{f^2 Q}\frac{1}{\phi^2}
\end{eqnarray}
which have to be smaller than one during the slow-roll epoch. Number of e-folding is another important parameter that will be used in perturbation analysis:
\begin{eqnarray}\label{Number of E-folding}
N=\int_{t_1}^{t} H dt=\frac{fQ}{8 M_p^2(1-f)}(\phi^2_2-\phi^2_1)
\end{eqnarray}  
where $\phi_1<\phi_2$ {(Eq.(\ref{scale}) has been used.)}. Now we are ready to introduce the perturbation parameters which are used to compare the model with the results of observation. Scalar power spectrum which can be computed  by data analysis of the CMB at pivot scale is presented for our model in high dissipative  regime :
\begin{eqnarray}\label{Scalar power-spectrum1}
\Delta_R^2=\frac{3\sqrt{3}Q^{\frac{5}{2}}}{8\pi^{\frac{3}{2}}M_p^2}[\frac{3M_p^2}{2C_{\gamma}}]^{\frac{1}{4}}\frac{(fA)^{\frac{9}{4}}}{(1-f)^{\frac{3}{4}}}[\frac{fQ}{24 A(1-f)}]^{\frac{3(3f-2)}{4f}}\phi^{\frac{3(3f-2)}{2f}}.
\end{eqnarray}  
{We have used Eqs.(\ref{Scalar power-spectrum}, \ref{Potential},\ref{slow-roll}) and the definition of   thermal bath temperature \cite{Herrera:2011zz}:
\begin{eqnarray}
T=[-\frac{M_p^2 \Upsilon\dot{H}}{2C_{\gamma}H Q}]^{\frac{1}{4}}.
\end{eqnarray}
}
Tensor perturbation power-spectrum is not modified in the context of warm inflation model. {Using Eqs.(\ref{Number of E-folding},\ref{Scalar power-spectrum1}) and the standard form of tensor power spectrum:}
\begin{eqnarray}
\Delta_T=\frac{2 H^2}{\pi^2 M_p^2},
\end{eqnarray}  
we can find the tensor-to-scalar ratio of our model:
\begin{eqnarray}\label{Tensor-to-scalar}
r=\frac{16(1-f)^{\frac{3}{4}}}{3\sqrt{3\pi}Q^{\frac{5}{2}}}(\frac{2C_{\gamma}}{3M_p^2 fA})^{\frac{1}{4}}[\frac{f(N-1)+1}{fA}]^{-\frac{f+2}{4f}}.
\end{eqnarray}
On the other hands, the spectral index $n_s$ is presented by 
\begin{eqnarray}\label{spectral index}
n_s-1=\frac{d\ln \Delta _R}{d\ln k}=-\frac{3(2-3f)}{4[f(N-1)+1]}
\end{eqnarray}
which leads to Harrison-Zeldovich spectrum with $f=\frac{2}{3}$ and red tilted cases where $f<\frac{2}{3}$. These are two perturbation parameters (\ref{Tensor-to-scalar},\ref{spectral index}) which are used to compare the inflation model with the results of observation \cite{Aghanim:2019ame}. We will compare them with the CMB data in next section.

\begin{figure}
    \includegraphics[width=.52\textwidth]{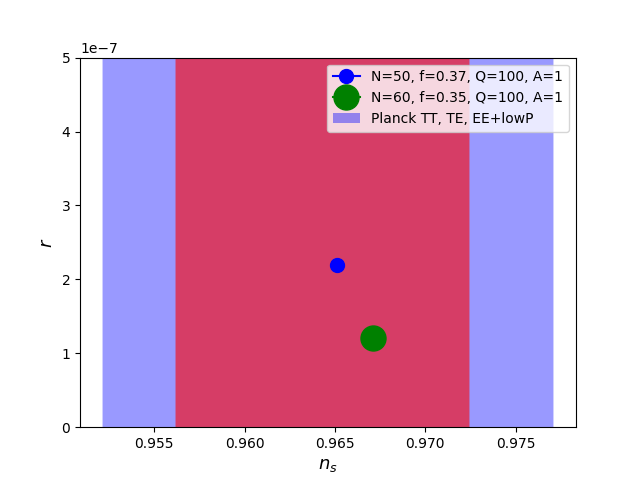}\hfill
    \includegraphics[width=.52\textwidth]{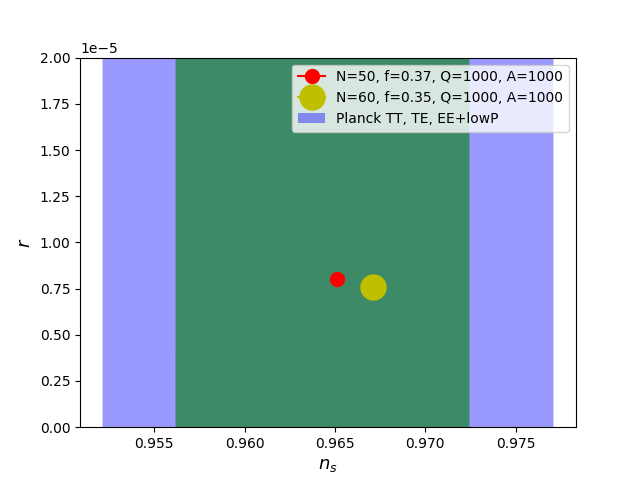}\hfill
    \caption{In these two pictures, we compare the theoretical results of our model with Planck observational data.}\label{Swampland1}
\end{figure}

\section{Constraints from quantum gravity, comparison with observation}
In this section, we present a region of free parameters phase space of the model which is compatible with observational data as well as quantum gravity. It was speculated that a scalar field theory that is used to study the early time accelerated expansion of the Universe is in the landscape if:
\begin{eqnarray}\label{sw1}
\frac{\Delta\phi}{M_p}\leq c_1,
\end{eqnarray}
and
\begin{eqnarray}\label{sw2}
\frac{V'}{V}>\frac{c_2}{M_p}~~~~~or~~~~~\frac{V''}{V}<-\frac{c_3}{M_p},
\end{eqnarray}
otherwise, it is in the swampland of string theory. $c_i$ is a constant of order one. The first inequality is distance conjecture and the second one is de Sitter conjecture \cite{Ooguri:2006in,Obied:2018sgi, Brennan:2017rbf, Agrawal:2018own,Bedroya:2019snp}. 
It was discussed in the literature that the high dissipative regime of warm inflation is in the landscape because of the modification of slow-roll parameters and tensor-to-scalar ratio in the context of warm inflation \cite{Das:2018rpg, Motaharfar:2018zyb, Kamali:2019hgv}. 
The main point is finding a region of free parameters phase space where the model is in landscape (generally high dissipative regime of warm inflation models) and compatible with observation data which is the main goal of this section. Another important conjecture is the Trans-Planckian censorship conjecture (TCC) \citep{Bedroya:2019snp} which forbids the horizon crossing of the sub-Planckian perturbation modes during the accelerated expansion. It constrains the energy scale of inflation and tensor-to-scalar ratio \cite{Bedroya:2019tba}:

\begin{eqnarray}\label{TCC1}
V^{\frac{1}{4}}<10^{10} GeV
\end{eqnarray} 

\begin{eqnarray}\label{TCC2}
r<10^{-30}
\end{eqnarray} 
We have found a limited region of the model parameters in which the model is compatible with limitations (\ref{TCC1},\ref{TCC2}).
In Fig. (\ref{Swampland1}), we have shown $1\sigma$ and $2\sigma$ confidence level of $r-n_s$ parameters which were borrowed from Planck results.  There are two points in these plots that are presented by our theoretical results for the number of e-folds $N=60$ and $N=50$ which are in $1\sigma$ confidence level of the $r-n_s$  Planck observational data. Therefore, our theoretical results agree with the observational data as well as swampland conjecture. 

\begin{figure}
    \includegraphics[width=.52\textwidth]{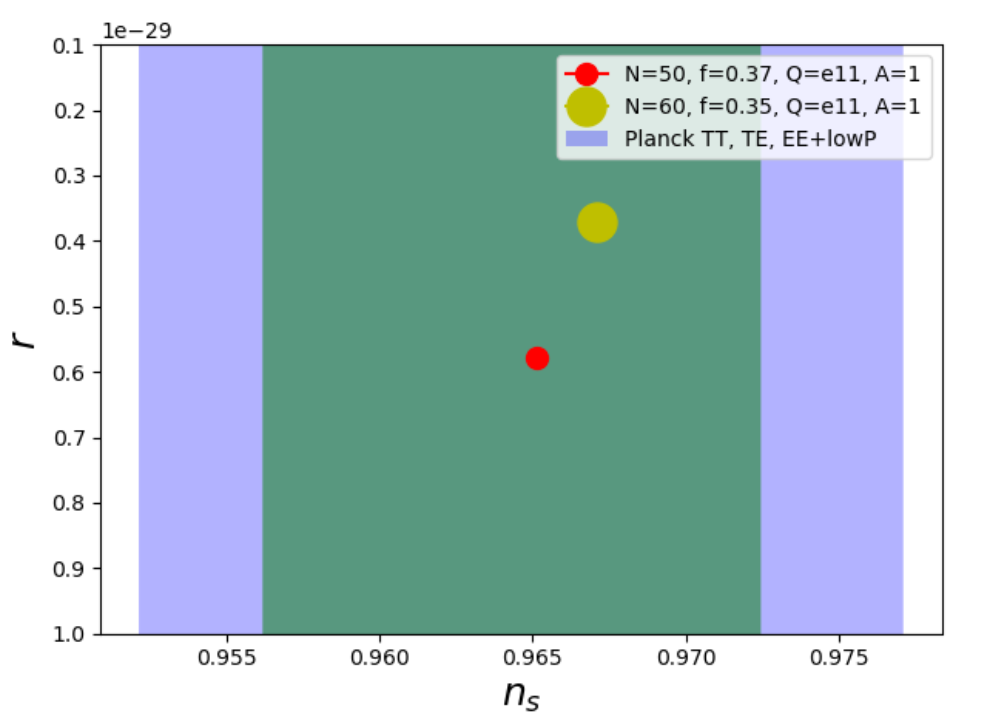}\hfill
    \caption{Extremely Large values of the $Q$ parameter are used to find the model compatible with observation, swampland, and TCC. }\label{TCC-Pic1}
\end{figure}

From Eqs.(\ref{slow-roll}, \ref{Tensor-to-scalar},\ref{spectral index}), It is obvious that the large values of the dissipation parameter $Q$ are useful. A limited region of constant $f$ ($\frac{1}{4}<f<\frac{2}{3}$) is used to satisfy the observation as well as swampland constraints (see Fig.(\ref{Swampland1})). On the other hand in Fig.(\ref{TCC-Pic1},
) we have used larger (limited) values of $Q$  to find the results compatible with observation, Swampland, and TCC. 
 {It seems one needs extremely huge values of $Q$ to find the model compatible with TCC. It means a lot of energy transfers to the gauge sector during inflation. In some cases, we can claim the model is compatible with TCC and observation but not a physical model. Generally, it is hard to find a physical model compatible with TCC in the context of intermediate warm inflation. }

\begin{figure}
	\includegraphics[width=.52\textwidth]{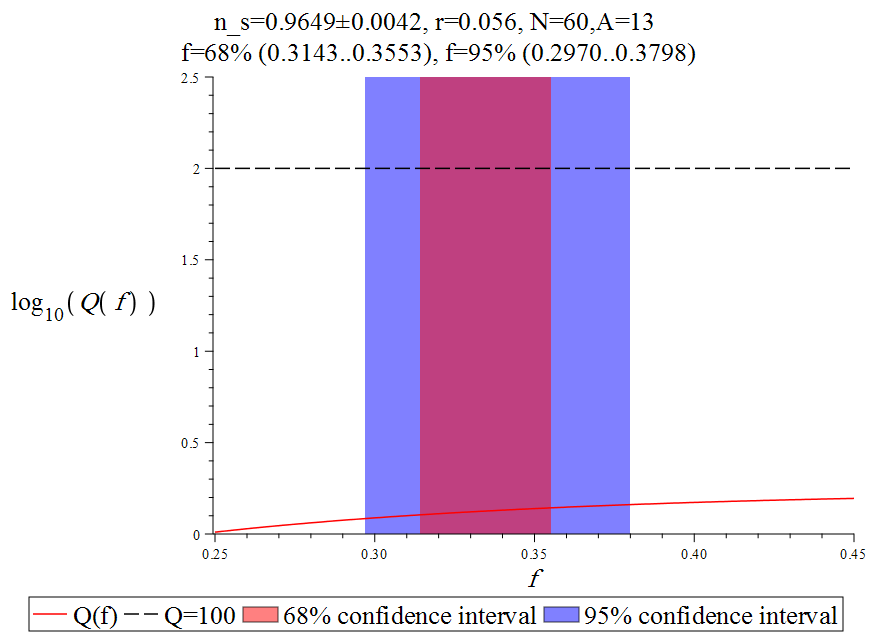}\hfill
	\caption{{In this figure, we fixed the number of e-folding and free parameter $A=13$. The colorful area between two red and black-dashed trajectories is compatible with just observation. The area upper than the black-dashed line is compatible with swampland and observation but not TCC.}}   \label{Observation-Swampland60}
\end{figure} 

\begin{figure}
	\includegraphics[width=.58\textwidth]{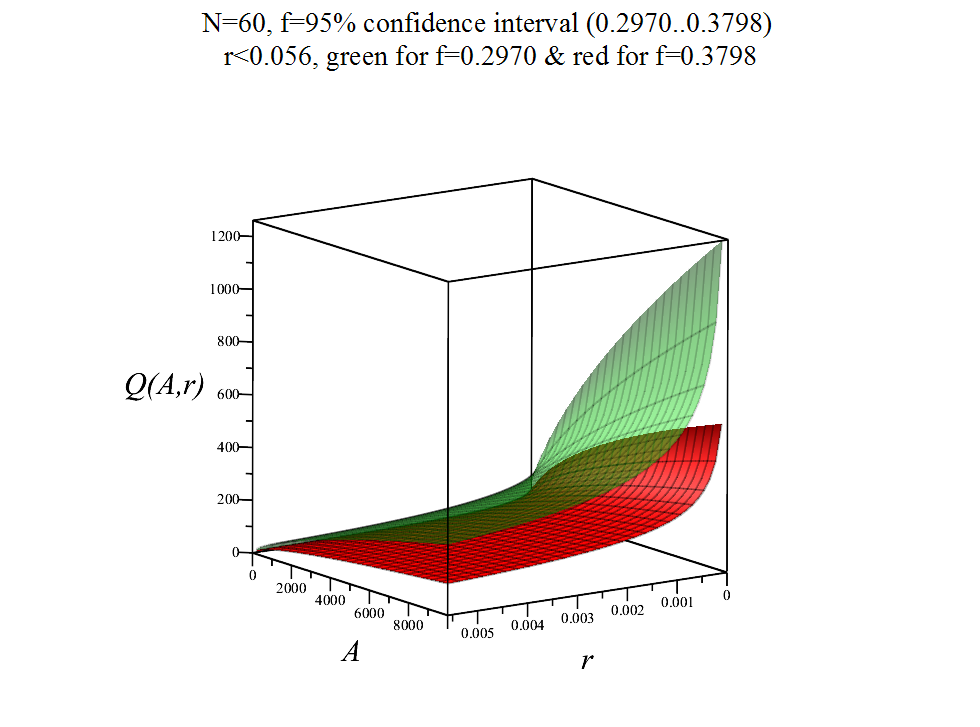}\hfill
	\caption{{In this figure, we show that the model is compatible with observation (the volume between two colorful surfaces)  where $A$ is large enough even with the smaller values of dissipative parameter $Q$. Two free parameters $A$ and $Q$ have the same effects on the tensor-to-scalar ratio in this case. }}\label{Observation-Swampland-QfA}
\end{figure}

{ 
 In Fig.(\ref{Observation-Swampland60}), we  show three distinguished area of free parameters $(Q,f)$ where $A=13$ and $N=60$. The area between the $f$ axis and red trajectory is rolled out by observation as well as quantum gravity. The colorful area between red and black-dashed trajectories is a free parameters phase space area that the model is compatible with observation but in the swampland. For the colorful area of free parameters phase space, upper than black-dashed line, our model is compatible with observation and in the landscape. In Fig.(\ref{Observation-Swampland-QfA}) we changed the other free parameter $A$. The model is in landscape and compatible with the observational data for the volume between two colorful surfaces where $Q>100$ \cite{Motaharfar:2018zyb}. }

\begin{figure}
	\includegraphics[width=.52\textwidth]{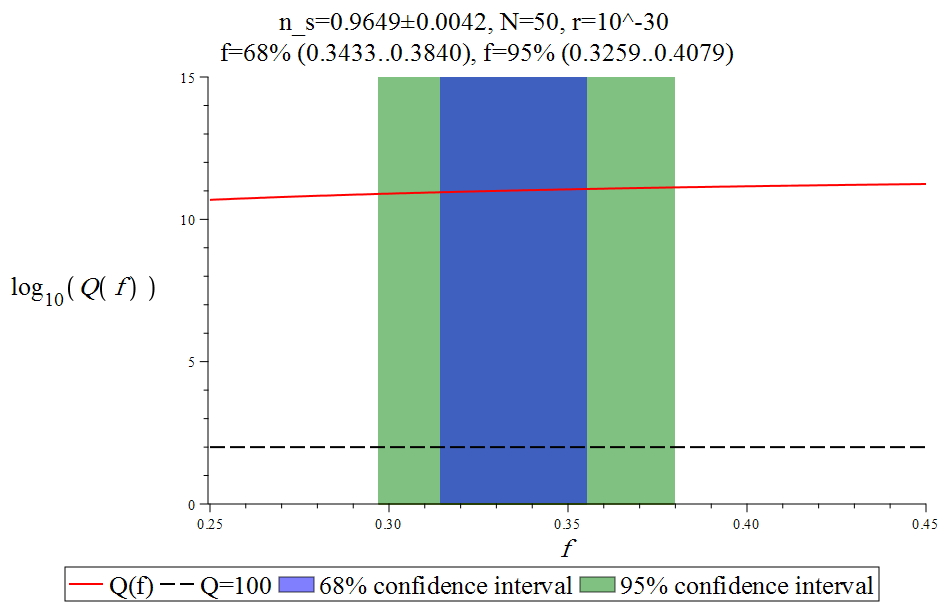}\hfill
	\caption{{There are two important regions in this plot. The first is between black-dashed and red lines where the model is in landscape and compatible with observation but not matched with TCC bound (\ref{TCC2}). The second region is upper than the red line which is also compatible with TCC, observation, and out of swampland. }}\label{Swampland-TCC60}
\end{figure}

\begin{figure}
	\includegraphics[width=.52\textwidth]{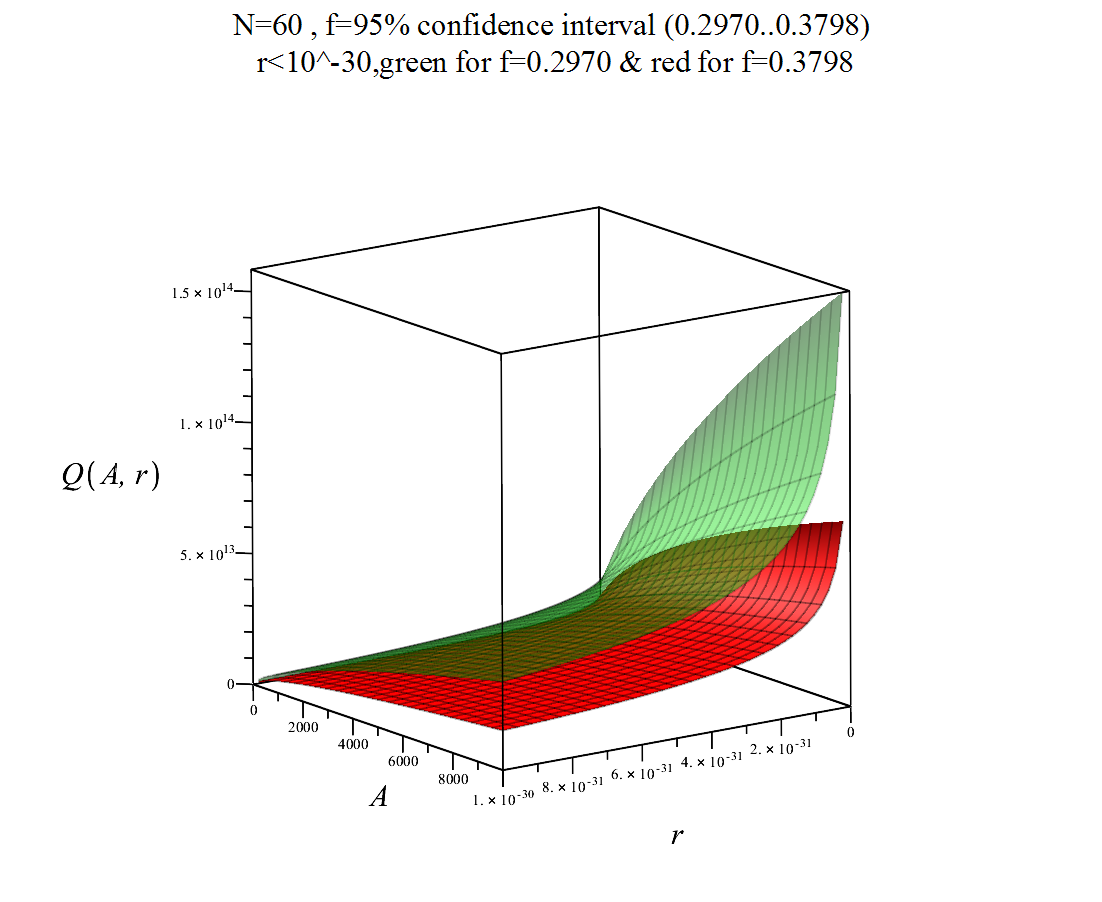}\hfill
	\caption{{There is a volume between two colorful surfaces which shows the model in the landscape, compatible with observation and TCC. }}\label{TCCQfA}
\end{figure}

{ In Fig.(\ref{Swampland-TCC60}), the area between black-dashed trajectory and red one introduces a free parameters phase space of the model which is in the landscape and compatible with observation but not matched with TCC condition (\ref{TCC2}). The model with free parameters upper than red line trajectory is in the landscape, compatible with observation and TCC condition (\ref{TCC2}). In Fig.(\ref{TCCQfA}), the volume between two colorful surfaces presents a region of the model parameters in the landscape which is compatible with observation and TCC. }

\section{Conclusion}
In this paper, we studied the intermediate solution of cosmic expansion in the context of warm pseudoscalar inflation. We have shown that the model is compatible with observational data for a region of the free parameters phase space. We also studied the constraints of quantum gravity on our model. It was shown that the model is compatible with swampland criteria and observation for a broad region of free parameters phase space. Our model is also compatible with TCC and observation for a limited region of free parameters phase space.{ Our analysis can be extended for the standard models of warm inflation \cite{Bastero-Gil:2017wwl,Arya:2018sgw}. The important point is that we need a high dissipative regime of warm inflation which is compatible with observation \cite{Motaharfar:2018zyb}. It would be our future works to find this regime for the standard warm inflation models \cite{Bastero-Gil:2017wwl,Arya:2018sgw}. }  

\begin{acknowledgements}:
We would like to thank Ehsan Barati and Reza Ghasemi for very useful comments on the draft. In memory of John D. Barrow and Shahriar Bayegan.  
\subsubsection{Shortcomings of Hot Big Bang theory}

\subsection{Standard model of cosmology}
Here we describe the dynamics of an expanding universe and also investigate the situation of the cosmos in the presence of different components of matter\cite{Wald}.  
\end{acknowledgements}
 \bibliographystyle{apsrev4-1}
  \bibliography{ref}
\end{document}